\documentclass[12pt]{article}\textheight 24.5cm \textwidth 17cm\voffset=-1.in\hoffset= - 1.in
\makeatletter
\@addtoreset{equation}{section}

\makeatother
\usepackage[utf8]{inputenc}
\usepackage[english]{babel}
\usepackage[T2A]{fontenc}
\usepackage{amsmath}
\usepackage{slashed}
\usepackage{graphicx}
\usepackage{amsfonts}
\usepackage{amssymb}
\usepackage{psfrag}
\usepackage{hhline}
\usepackage{cite}
\usepackage{upgreek}
\usepackage{epsfig,amsfonts}

\author{ A. Belavin\footnote{ belavin@itp.ac.ru}$\,\,^{1,2,3}$, L. Spodyneiko\footnote{lionspo@itp.ac.ru}$\,\,^{1}$
\vspace*{10pt}\\[\medskipamount]
$^1$~\parbox[t]{0.88\textwidth}{\normalsize\it\raggedright
L.D.Landau Institute for Theoretical Physics,
142432 Chernogolovka, Russia}
\vspace*{10pt}\\[\medskipamount]
$^2$~\parbox[t]{0.88\textwidth}{\normalsize\it\raggedright
Moscow Institute of Physics and Technology, 141700 Dolgoprudny, Russia}
\vspace*{10pt}\\[\medskipamount]
$^3$~\parbox[t]{0.88\textwidth}{\normalsize\it\raggedright
Institute for Information Transmission Problems, 127994 Moscow, Russia}
}
\date{}
\title{\bf $N=2$ superconformal algebra in NSR string and Gepner approach to space-time supersymmetry \\in ten dimensions}
\begin{document}
\maketitle
\begin{abstract}

The fermionic NSR string possesses a hidden $N=2$ superconformal algebra on the world-sheet. In this work, we show how to use an isomorphism of this algebra, the so-called spectral flow, for construction of a subspace of physical states of the string, on which space-time supersymmetry acts. This construction is an alternative to the GSO-projection in string theory.

\end{abstract}
\vspace*{10pt}
\begin{center}{\it The paper is based on lectures given at Independent University of Moscow in spring  2015.}
\end{center}
\vspace*{10pt}
\section{Introduction}
Superstring theory \cite{Green:1987sp, Polchinski:1998rq, Becker:2007zj, Blumenhagen:2013fgp} plays an important role in modern theoretical and mathematical physics. An essential feature of this theory is the space-time supersymmetry, which is proposed to be a natural mechanism to solve the hierarchy problem.

 It was shown by Gliozzi, Sherk and Olive\cite{Gliozzi:1976jf, Gliozzi:1976qd} that space-time supersymmetry in the fermionic Neveu-Schwartz-Ramond string\cite{Neveu:1971rx,Ramond:1971gb}  appears after the special projection of physical states with use of the world-sheet fermionic number operator. They have shown that after the projection  the amount of physical states in bosonic (NS) and  fermionic (R) sectors are the same for each level of a given mass. The odd super-Poincare operators in the covariant approach were built by Friedan, Shenker and Martinec \cite{Friedan:1985ey,Friedan:1985ge}, and by Knizhnik\cite{Knizhnik:1985ke}. In these works, the authors used "spin" field in the matter sector, as well as,  bosonization of ghost sector for the construction of massless fermionic vertex. The limit of this vertex at zero momentum gives covariant space-time supersymmetry generator of the NSR string. In addition, the GSO-project is necessary in order for the space-time supersymmetry generators to be well defined.

 In \cite{Gepner:1987qi},\cite{Banks:1987cy} it was shown that the condition for supersymmetry in space-time after compactification of ten-dimensional strings to four-dimensional Minkowski space is the $ N = 2 $ superconformal symmetry of the theory in six compact dimensions. Gepner\cite{Gepner:1987qi} explained how the operator of the so called spectral flow which maps NS-sector in R-sector and vice versa, can be used to construct the space-time SUSY generator.

Meanwhile, the original $ d = 10 $ NSR string itself has a hidden $ N = 2 $ superconformal symmetry on the world-sheet. In this article, we show how to use the operator of the spectral flow $ U $, which transfer NS- and R- sectors into each other, to restrict the local physical field space by the requirement of their locality with respect to the operator $ U $. Thereafter, we determine the action of space-time supersymmetry on this subspace.

The requirement of locality is equivalent to the GSO-projection and the construction itself with use of the full $ N = 2 $ superconformal symmetry algebra of the NSR string in the BRST approach leads to the same (up to the action of the picture-changing operator) supercharge formula as for $ d = 10 $ superstring in works \cite{Friedan:1985ey, Knizhnik:1985ke}. However, such a formulation, which is absent in the standard textbooks like \cite{Green:1987sp, Polchinski:1998rq, Becker:2007zj, Blumenhagen:2013fgp}, is useful. First, it clarifies the procedure of achieving space-time SUSY in ten dimensions. Second, we naturally arrive to the Gepner hypothesis on the necessity of $N=2$ superconformal symmetry to hold for six compactified dimensions. Compactification must preserve the $N=2$ superconformal symmetry on the world-sheet, which is present in the ten-dimensional string.

\section{Conformal Field Theory}
Any version of string theory is a special case of two-dimensional conformal field theory \cite{BPZ}. In this section we briefly remind basic ideas of CFT including local fields, operator product expansions (OPEs), holomorphic currents and action of Fourier modes on local fields. There are different ways to describe strings like Hilbert space formalism, path integral formulation e.t.c.. In this paper we use language of local fields and their OPEs in spirit of \cite{Friedan:1985ge}.

Two-dimensional conformal field theory is a field theory with the vanishing of the trace of the energy-momentum tensor
\begin{align}
T^a_a  = 0.
\end{align}
We are interested in the local fields of this theory. An important characteristic of the fields is their conformal dimensions $ \Delta, \overline \Delta $ defined as
\begin{align}
\Phi(z,\bar z ) = \lambda^\Delta \lambda^{\overline \Delta} \Phi(\lambda z, \lambda \bar z).
\end{align}

In conformal field theory, it is convenient to use the operator product expansion. Operator product expansion (OPE) is a decomposition of  two fields in the correlation functions $ \langle \dots \Phi_1 (z, \bar z) \Phi_2 (w, \bar w) \dots \rangle $ in a combination of local fields, when operators are close to each other $ z \rightarrow w $. For example, the OPE of the energy-momentum tensor with itself reads
\begin{align}\label{tt ope}
T(z)T(w)&\sim \frac c {2(z-w)^4} + \frac 2 {(z-w)^2} T(w) + \frac 1 {z-w} \partial T(w)+\text{regular terms}.
\end{align}
Here and below we omit the fact that this expression is understood as part of a correlation function. We will also write  only the singular part of the expansions.

The variation of the fields under infinitesimal transformations $ z \rightarrow z + \varepsilon (z) $ is given by
\begin{align}
\delta_\varepsilon \Phi = \int dz \, \varepsilon(z)T(z) \Phi(0,0).
\end{align}
Choosing $ \varepsilon (z) = z ^ {n + 1} $, we can define the operators
\begin{align}
L_n = \int dz \, z^{n+1} T(z),
\end{align}
where the integral is taken around the point $ z = 0 $ and the operators $L_n$ act on the field at this point. Using the operator product expansion (\ref {tt ope}), it can be shown that the commutator of these operators is
\begin{align}\label{lnlm}
[L_m,L_n] &= (m-n)L_{m+n} +\frac c {12} (m^3-m)\delta_{m,-n}.
\end{align}
Similarly, there is a correspondence between other holomorphic currents and symmetry algebras of string theory.
\section{NSR string}
The action of the matter part of the free ten-dimensional NSR-string \cite{Neveu:1971rx, Ramond:1971gb} is
\begin{align}
S_m = \int d^2 z \, \left[ \partial X^\mu \overline \partial X_\mu + \psi_\mu \overline\partial \psi^\mu + \widetilde \psi_\mu \partial \widetilde\psi^\mu\right],
\end{align}
where $ \mu = 0, \dots, 9 $ and $ X_\mu, \psi_\mu, \widetilde \psi_\mu $ are  matter fields. Below, we are interested only in the holomorphic part of the theory. These fields have the following OPE's
\begin{align}
\begin{split}
X_\mu(z) X_\nu(0) &\sim- \eta_{\mu\nu} \ln z,\\  \psi_\mu(z) \psi_\nu(0) &\sim \frac {\eta_{\mu\nu}} z,
\end{split}
\end{align}
where the metric $ \eta _ {\mu \nu} = \text {diag} (- 1,1, \dots, 1) $. All other OPE's are regular. The fields $ \psi_\mu $  have the following monodromy around the vertex operator at the point $ z = 0 $
\begin{align}\label{monodromy}
\psi_\mu(e^{2\pi i} z ) = e^{2\pi i\nu } \psi_\mu(z).
\end{align}
When $ \nu = 1/2 $, the vertex operator at the point $ z = 0 $ is said to be in the R-sector, and when  $ \nu = 0 $ it is said to be in the NS-sector.

As is well known, this theory has  $ N = 1 $ superconformal symmetry. The corresponding energy-momentum tensor and supercurrent are
\begin{align}
\begin{split}
T^m &= -\frac 1 2 \partial X^\mu \partial X_\mu - \frac 1 2 \psi^\mu \partial \psi_\mu ,\\
G^m &= i \psi^\mu \partial X_\mu .
\end{split}
\end{align}
Below the letter $ m $ in expressions like $ T ^ m $ denotes the matter sector.

The currents $ T ^ m (z) $ and $ G ^ m (z) $ satisfy the  $ N = 1 $ superconformal algebra OPE's
\begin{align}
\begin{split}
T(z)T(0)&\sim \frac c {2z^4} + \frac 2 {z^2} T(0) + \frac 1 z \partial T(0),\\
T(z)G(0)&\sim \frac 3 {2z^2} G(0) + \frac 1 z \partial G(0),\\
G(z)G(0)&\sim \frac {2c}{3z^3} + \frac 2 z T(0),
\end{split}
\end{align}
with  central charge $ c = c_m = $ 15. If one expands the currents in Fourier modes $ T (z) = \sum _ {n \in \mathbb Z} L_n z ^ {- n-2} $, $ G (z) = \sum_ {r \in \mathbb Z + \nu + 1/2} G_r z ^ {- r-3/2} $, the commutation relations take the form

\begin{align}\label{vir}
[L_m,L_n] &= (m-n)L_{m+n} +\frac c {12} (m^3-m)\delta_{m,-n},\\
\{G_r,G_s\} &= 2 L_{r+s} + \frac c {12} (4 r^2 -1), \delta_{r,-s}\\
[L_m,G_r] &= \frac{m-2r} 2 G_{m+r},
\end{align}
where $ n, m \in \mathbb Z $, $ r, s \in \mathbb Z + \nu + 1/2 $. The modes of $ G (z) $ are expanded in integer modes in the R-sector and in half-integer modes in the NS-sector.

These commutation relations lead to an important inequality. For each field $ \Phi $ in the R-sector we have
\begin{align}\label{ramond ineq}
\Big|G_0|\Phi\rangle\Big|^2 = \frac 1 2 \langle \Phi |\{G_0,G_0\}|\Phi \rangle  =  \langle \Phi |\left(L_0 - \frac c {24}\right)|\Phi\rangle =\left(\Delta - \frac c {24}\right)\langle \Phi |\Phi \rangle\ge0.
\end{align}
It follows that in a unitary  theory the dimensions of all fields in the R-sector satisfy the bound $ \Delta \ge \frac c {24} $. Equality holds when the field satisfies $ G_0 | \Phi \rangle = 0 $. The field with conformal dimension $ \Delta = \frac c {24} $ and zero space-time  momentum is called Ramond vacuum.

In the matter sector there are 32 Ramond vacua $ S_ \alpha (z) $, with $ \alpha = 1, \dots,32 $. These 32 vacua transform as  a 32-component spinor. Namely
\begin{align}
\psi^\mu(z) S_\alpha(0)\sim \frac 1 {\sqrt{2z}} \Gamma^\mu_{\alpha\beta} S_\beta (0),
\end{align}
where $ \Gamma ^ \mu $ are gamma matrices of size $ 32 \times 32 $. Note that the square root of $z$ on the right-hand side leads to the minus sign after translation of the field $ \psi_ \mu (z) $ around zero, as it should be in the R-sector according to the formula (\ref {monodromy}).

In string theory the currents $ T ^ m (z), G ^ m (z) $ are constraints. Using these currents, one can construct the BRST-charge. It should include two fermionic ghosts $ b, c $, corresponding to constraints generated by the current $ T ^ m (z) $, and two bosonic ghosts $ \beta, \gamma $,  corresponding to $ G ^ m (z) $. The explicit expression for the BRST-charge is

\begin{align}\label{BRST}
Q_B&=\int dz \left[c T^m + \gamma G^m + \frac 1 2 \left (cT^{gh} + \gamma G^{gh}\right)\right].
\end{align}
The physical states $\Phi$  are defined as cohomologies of the BRST-charge
 \begin{align}\label{ph st def}
 \begin{split}
 Q_B \Phi &= 0,\\
 \Phi&\simeq \Phi + Q_B\Psi.
 \end{split}
 \end{align}
 The last formula means that fields that differ by a  BRST-exact term $ Q_B \Psi $ for any $ \Psi $ are physically equivalent. The so defined physical states have a positive norm and their spectrum coincides with the one of the superstring in the light-cone gauge.

The ghost fields have the OPE's
\begin{align}\label{ghost ope}
\begin{split}
\beta(z) \gamma(0)&\sim - \frac 1 z, \\ b(z) c(0)&\sim\frac 1 z.
\end{split}
\end{align}
The ghost action is
\begin{align}
S_{gh} = \int d^2z \left[b\overline \partial c + \beta \overline \partial \gamma+h.c.\right].
\end{align}
This action has an  $N=1$ superconformal symmetry with the currents
\begin{align}
\begin{split}
T^{gh} &= - \partial b c -2 b \partial c - \frac 1 2 \partial \beta \gamma - \frac 3 2 \beta \partial \gamma,\\
G^{gh} &= \partial \beta c + \frac 3 2 \beta \partial c - 2 b\gamma.
\end{split}
\end{align}
They satisfy the relations of the superconformal algebra with central charge $ c_ {gh} = - $ 15.

The monodromy of the fields $ \beta, \gamma $ also depends on the sector. In order for the BRST-charge to be well-defined, the integrand in (\ref {BRST}) must be periodic around zero. Therefore, the fields $ \beta, \gamma $ must have the same monodromy as $ \psi_ \mu $, i.e. the ghost and  material components of vertex operator at the point $ z = 0 $ must be in the same sector

\begin{align}
\begin{split}
\psi_\mu(e^{2\pi i} z ) = e^{2\pi i\nu } \psi_\mu(z),\\
\beta(e^{2\pi i} z ) = e^{2\pi i\nu } \beta(z),\\
\gamma(e^{2\pi i} z ) = e^{2\pi i\nu } \gamma(z),
\end{split}
\end{align}
where $\nu= 1/2 $ in R-sector, $\nu= 0$ in NS-sector.

$\beta- \gamma$  system has vacua $ V_q (z) $ parameterized by a half-integer $ q $ called the number of the picture \cite{Verlinde:1988tx}, and the space of states generated by $ \psi_\mu, \partial X_\mu, \beta, \gamma, b, c $  out of the vacuum $ V_q $, is called the picture. These spaces are not isomorphic to each other for different values of $ q $ without physical constraints. The vacua $ V_q $ are determined by the  conditions
\begin{align}\label{picture}
\begin{split}
\beta(z) V_q(0) &\sim O(z^q),\\
\gamma(z) V_q(0) &\sim O(z^{-q}).
\end{split}
\end{align}
Note, that from these formulas follows  that the translation of  $ \beta, \gamma $ around the origin produces a phase $ e ^ {2 \pi iq} $. Therefore,  $ q $ must be an integer in the NS-sector and half-integer in the R-sector.

One can build physical states in  different pictures. It is known \cite {Verlinde:1988tx} that the physical states (BRST-cohomologies) in the different pictures are isomorphic to each other. The isomorphism is given by the action of the so-called picture changing operator. It is convenient to choose the canonical pictures  $ q = -1 / 2 $ in the R-sector and $ q = -1 $ in the NS-sector.

The general form of a vertex in the NS-sector is
\begin{align}
P(\psi_\mu,\partial X_\mu,\beta,\gamma,b,c)V_q e^{ik_\mu X^\mu},
\end{align}
where $P$ is a polynomial of its arguments, and $k_\mu$ is a momentum. In the NS-sector $q$ must be an integer.

The general form of a vertex in the R-sector is
\begin{align}
P^\alpha(\psi_\mu,\partial X_\mu,\beta,\gamma,b,c) S_\alpha V_q e^{ik_\mu X^\mu},
\end{align}
where $P^\alpha$ is a polynomial of its arguments and it transforms as a 32-component spinor. In the R-sector $q$ is half-integer.

Two important examples of vertex operators are  massless bosons and fermions in the pictures $-1$ and $-1/2$
\begin{align}
\begin{split}
V_{NS} &= \xi_\mu \psi^\mu V_{-1} e^{ik_\mu X^\mu}\\
V_{R} &= u^\alpha S_\alpha V_{-1/2} e^{ik_\mu X^\mu}
\end{split}
\end{align}
where $u_\alpha$, $\xi_\mu$ are polarizations.
\section{The $N=2$ superconformal algebra}\label{n=2}
In this section we define the  $ N = 2 $ superconformal algebra and its isomorphism, the so-called spectral flow, which later will play a decisive role.

The $N=2$ superconformal algebra consists of the currents  $T(z),G^{\pm}(z),J(z)$. The current $T(z)$ is a energy-momentum tensor with dimension 2, $G^\pm(z)$ are two supercurrents with dimension $3/2$ and  $J(z)$ is a $U(1)$-current with dimension 1, which corresponds to the R-symmetry of the $N=2$ superalgebra. These currents have OPE's
\begin{align}
\begin{split}
T(z) G^{\pm}(0) &\sim \frac 3 {2z^2} G^\pm(0) + \frac 1 z \partial G^{\pm}(0),\\
T(z) J(0)&\sim \frac 1 {z^2}J(0)+ \frac 1 z \partial J(0),\\
G^+(z)G^-(0)&\sim \frac {2c}{3z^3} + \frac 2 {z^2} J(0) + \frac 2 z T (0)+ \frac 1 z \partial J(0),\\
G^\pm(z)G^\pm(0)&\sim 0,\\
J(z)G^\pm(0) &\sim \pm \frac 1 z G^\pm(0),\\
J(z)J(0)&\sim \frac c {3z^2}.
\end{split}
\end{align}
This algebra has an $N=1$ subalgebra generated by the currents $T(z), G(z)= (G^++G^-)/\sqrt 2$.

After Fourier mode expansion
\begin{align}
\begin{split}
G^{\pm}(z) &= \sum_{r\in \mathbb Z + \frac 1 2 \pm \nu} G^{\pm}_r z^{-r - 3/2},\\
T(z)&=\sum_{n\in \mathbb Z} L_n z^{-n-2},\\
J(z) &= \sum_{n\in \mathbb Z} J_n z^{-n-1},
\end{split}
\end{align}
the commutation relations become
\begin{align}
\begin{split}
[L_m,G_r^\pm] &= \left(\frac m 2 - r \right)G^\pm_{m+r},\\
[L_m,J_n] &= - n J_{m+n},\\
\{G^+_r,G^-_s\}&= 2L_{r+s}+ (r-s)J_{r+s}+ \frac c 3 \left(r^2 - \frac 1 4\right)\delta_{r+s,0},\\
\{G^\pm_r,G^\pm_s\}&=0,\\
[J_n,G^{\pm}_r]&=\pm G^\pm_{r+n},\\
[J_m,J_n]&=\frac c 3 m \delta_{m+n,0}.
\end{split}
\end{align}
The value of  $\nu$ depends on the monodromy  of $G^{\pm}(z)$
\begin{align}
G^{\pm}(e^{2\pi i } z ) = e^{\pm 2\pi i \nu}G^{\pm}(z),
\end{align}
where $\nu$ can have an arbitrary real value. We will be mainly interested in the particular cases $\nu=0$ and $\nu = \frac 1 2$, corresponding to the NS- and R- sectors.

The $N=2$ superconformal algebra possesses an isomorphism  \cite{Schwimmer:1986mf}

\begin{align}
\begin{split}\label{automorph}
L'_n &= L_n + \eta J_n + \frac 1 6 \eta^2 c \delta_{n,0},\\
J'_n &= J_n + \frac c 3 \eta \delta_{n,0},\\
(G^{\pm}_r)' &= G^\pm_{r\pm \eta}.
\end{split}
\end{align}
The action of this isomorphism on the representations of the superconformal algebra can be realised\cite{Gepner:1987qi, Gepner:1989gr}  in terms of the bosonic scalar field  $\varphi(z)$. One can bosonize the $U(1)$-current $J(z)$
\begin{align}\label{j boson}
\begin{split}
J(z)& = \partial \varphi(z),\\
\varphi(z)\varphi(0)&\sim \frac c 3 \ln z.
\end{split}
\end{align}
Note that the boson $ \varphi (z) $ depends on the realization of the generators of the $ N = 2 $ superconformal algebra in terms of the fields of the theory in which this algebra acts. It will be shown that   in the matter and ghost sectors there is an $ N = 2 $ superconformal algebra and we will give expressions relating the bosonization of the $ U (1) $-currents in each of the sectors with the scalar fields of these sectors.

For an arbitrary field $V$, with charge $q$ under the current $J$, we can isolate the charged part
\begin{align}
V = \hat V e^{i\frac {3q} c \phi},
\end{align}
where $\hat V$ is neutral under $J(z)$. This procedure for $G^\pm(z)$ gives
\begin{align}
G^\pm = \hat G^\pm e^{\pm\frac 3 c \phi }.
\end{align}
For every field $V$ in a representation of the superconformal algebra we can construct a field twisted by  $\eta$
\begin{align}
V_\eta =V e^{\eta \phi }=\hat V e^{(\frac {3q} c + \eta) \phi }.
\end{align}
By straightforward computation one can show that charge of the twisted field is
\begin{align}
q' = q+ \frac c 3 \eta.
\end{align}
Also, if, for example, the original field has an OPE with the $ G ^ {\pm} $ in integer powers  $ z ^ n $, then the field  $ V_ \eta $ will have OPE with  $ G ^ {\pm} $ in powers  $ z ^ {n \pm \eta} $. The additional power arises from the OPE of $ \exp (\eta \phi) $ with $ \exp (\pm \frac 3 c \phi) $. The conformal dimension of the field  $ V_ \eta $ is

\begin{align}
\Delta' = \Delta+ \frac c {6} (\frac {3q}c + \eta )^2 - \frac {3q^2}{2c} = \Delta + \eta q + \frac 1 6 \eta^2 c.
\end{align}
From these formulas it is easy to see that multiplication of the fields on the vertex $ \exp {\eta \phi} $ corresponds to the isomorphism (\ref{automorph}). The action of this isomorphism will be denoted as the $ U_{\eta} $.

The physical states of the NS-sector are the space-time bosons, and the states of the R-sector are fermions. The fact that the spectral flow with $ \eta = \pm 1/2 $ translates NS- into  R- sector and backwards suggests that the corresponding vertex operator $ \exp (\eta \phi) $ is the supercharge or at least its component. In what follows, we will show that it is true.

In the $N=2$ superconformal algebra there is a restriction on the dimension $\Delta$ of the field with $U(1)$-charge $q$. For a field $\Phi$ in the NS-sector one can write
\begin{align}\label{chiral ineq}
 \left|G^\mp_{-1/2}|\Phi\rangle\right|^2+ \left|G^\pm_{1/2}|\Phi\rangle\right|^2 = \langle \Phi| \{G^\pm_{1/2},G^{\mp}_{-1/2}\} |\Phi\rangle =  (2\Delta \pm q) \langle \Phi|\Phi\rangle\ge0.
\end{align}
From this equation it follows that in a unitary theory in the NS-sector there is an inequality
\begin{align}
2\Delta \ge |q|.
\end{align}
The fields with $2\Delta = q$ or $2\Delta = - q$ are called correspondingly chiral or antichiral primary fields \footnote{We will sometimes call the field chiral meaning that it is chiral or antichiral primary field.}. As follows from (\ref{chiral ineq}), chiral primary field $\Phi$ satisfies
\begin{align}\label{chiral}
G^-_{1/2} \Phi = G^+_{-1/2} \Phi = 0.
\end{align}
Using this formula, the relations of the superconformal algebra and the restriction on the dimensions $2\Delta \ge |q|$, one can show that  $\Phi$ is annihilated by positive modes of the currents  $G^{\pm}(z)$, $T(z)$, $J(z)$. Along with (\ref{chiral}), it gives
\begin{equation}\label{chiral mode def}
\begin{aligned}
L_n \Phi &= J_n \Phi = 0, & &n>0,\\
G^+_r \Phi&= 0, & &r\ge-\frac 1 2,\\
G^-_r \Phi&=0, & &r>0.
\end{aligned}
\end{equation}
There are similar relations for the antichiral primary fields with  $G^+$ and $G^-$ exchanged.

 Since the $N=2$ superconformal algebra has the $N=1$ subalgebra, there is a restriction on the dimension $\Delta \ge \frac c {24}$ for the Ramond fields. It follows from the inequality (\ref{ramond ineq}). Moreover, the Ramond field $\Phi$ with the dimension  $\Delta = \frac c {24}$ satisfies
 \begin{align}
 G_0 \Phi = 0.
 \end{align}
 Because of the restriction  $\Delta \ge \frac c {24}$, $\Phi$ is annihilated by all the positive modes of the currents  $G^{\pm}(z)$, $T(z)$, $J(z)$. Using the commutation relations of the $N=2$ algebra, one can show that if the field is annihilated by $G_0 = (G^+_0+G^-_0)/\sqrt 2$, then it is annihilated by  $G^\pm_0$ separately. All this reads

\begin{equation}\label{ramond mod def}
\begin{aligned}
L_n \Phi &= J_n \Phi = 0,, & &n>0,\\
G^\pm_n \Phi&=0, & &n\ge0.
\end{aligned}
\end{equation}

It turns out, that the action of the spectral flow $U_{\pm 1/2 }$ transforms the  Ramond vacua and chiral fields into each other. Indeed, as it follows from  (\ref{automorph}) with $n=0$ the spectral flow acts as
\begin{align}\label{ch -> ram}
\begin{split}
U_{1/2} \left|{\Delta = \frac q 2\atop Q = q }\right\rangle_{NS} &=\left|{\Delta = \frac c {24}\atop Q=q - \frac c 6}\right\rangle_R,\\
U_{-1/2} \left|{\Delta = \frac q 2\atop Q = -q }\right\rangle_{NS} &=\left|{\Delta = \frac c {24}\atop Q=-q + \frac c 6}\right\rangle_R.
\end{split}
\end{align}
As it will be explained below, the vertex operators of the massless bosons are chiral fields. And the vertex operators of the massless fermions are the Ramond vacua. The relations  (\ref{ch -> ram}) mean that these states form a supermultiplet.
\section{The $N=2$ superconformal algebra in the NSR string}
As we mentioned earlier the matter and ghost sectors of  the NSR string in ten dimensions have the $ N = 2 $ superconformal symmetry. In this section we  describe them explicitly. Also we  give standard formulas for bosonization. After bosonization the $ U (1) $-currents will take a simple form.
\subsection{The matter sector}
It is convenient to choose another basis in the matter sector
\begin{align}
\begin{split}
\psi_0^\pm &= \frac 1 {\sqrt 2} (\pm\psi_0 +\psi^1 ),\\
\psi_k^\pm&= \frac 1 {\sqrt 2} (\psi^{2k} \pm i \psi^{2k+1}),\\
X_0^\pm &= \frac 1 {\sqrt 2} (\pm X_0 +X^1 ),\\
X_k^\pm&= \frac 1 {\sqrt 2} (X^{2k} \pm i X^{2k+1}).
\end{split}
\end{align}
The OPE's in this basis are
\begin{align}\label{psi +- ope}
\psi^+_a(z) \psi^-_b(0) &\sim \frac {\delta_{ab}}{z},\\
\partial X^+_a(z) \partial X^-_b(0) &\sim -\frac {\delta_{ab}}{z^2}.
\end{align}
The other OPE's are regular. The supercurrent reads
\begin{align}
G^m = i \psi^\mu \partial X_\mu = \sum_k i  \psi_k^+ \partial X_a^- +\sum_ki \psi^-_k \partial X^+_k,
\end{align}
One can show that the currents  $G^m_\pm$
\begin{align}
\begin{split}
G^m &= \frac 1 {\sqrt 2} (G^m_++G^m_-),\\
G^m_+ &= \sum_k i\sqrt{2} \psi_k^+ \partial X_k^-,\\
G^m_- &=\sum_k i\sqrt{2} \psi^-_k \partial X^+_k,
\end{split}
\end{align}
and the $U(1)$-current
\begin{align}
J^m = \sum_k \psi^+_k \psi^-_k,
\end{align}
together with  $T^m$ form the  $N=2$ superconformal algebra with central charge $c_m =15$.

The ten fermions $\psi_\mu$ can be realised in terms of five independent bosons $H_k$. These bosons have the OPE's
\begin{align}
H_a(z)H_b(0) &\sim - \delta_{ab}\ln z,
\end{align}
where $a,b=1,\dots,5$. One can show that fields
\begin{align}
\psi^\pm_k &= e^{\pm iH_k}
\end{align}
have the  OPE's (\ref{psi +- ope}). The energy-momentum tensor of the fermions reads
\begin{align}\label{hk t}
T_{\psi} = i\psi_\mu \partial \psi^\mu = -\frac 1 2 \sum_k \partial H_k \partial H_k.
\end{align}
The Ramond vacua in terms of the bosons  $H_k$ have a simple form
\begin{align}\label{spinors}
S_\alpha = e^{\sum_k i s_k H_k},
\end{align}
where $s_k = \pm 1/2 $. The number of different combinations of  $s_k$ is 32 and coincides with independent polarizations of the 32-component spinor. With the help of  (\ref{hk t}) one can find the dimension of these fields. It is  $5/8 = c_m/{24}$, as it should be for the Ramond vacua.

Finally, we will rewrite  the $U(1)$-current  in terms of  $H_k$. Using
\begin{align}
\psi_k^+\psi_k^- & = i \partial H_k, \quad \text{(no summation)}
\end{align}
we have
\begin{align}\label{matter u1}
J^m &= \partial H^m,
\end{align}
where
\begin{align}\label{matter u1}
H^m &= \sum_k {i H_k }.
\end{align}
Note, that the choice of the $N=2$ superconformal algebra is ambiguous. One can take another $U(1)$-current
\begin{align}
J^m = \psi_\mu
\Lambda_{\mu\nu}\psi_\nu,
\end{align}
where $\Lambda_{\mu\nu}$ is a nondegenerate antisymmetric matrix with eigenvalues $\pm 1$. The supercurrents  $G^m_\pm$ are the parts of $G^m$ with $U(1)$-charges $\pm1$  respectively. By a Lorentz transformation one can bring this current to a form  (\ref{matter u1}). As it would be clear later, different choices of $N=2 $ superconformal algebra in the matter sector correspond to different polarizations of 16-component supercharge  $Q_\alpha$.
\subsection{The ghost sector}
As it was noticed in \cite{Friedan:1985ge} the ghost sector has  the $N=2$ superconformal symmetry. We will show that the spectral flow in the ghost sector is a part of the total spectral flow which realizes the transformation between bosons and fermions.

The supercurrents $G^{gh}_\pm$
\begin{align}
\begin{split}
G^{gh} &= \frac 1 {\sqrt 2 } (G^{gh}_+ + G^{gh}_-),\\
G^{gh}_+&= \sqrt 2 \partial \beta c + \frac 3 {\sqrt 2 } \beta \partial c,\\
G^{gh}_- &= -2\sqrt 2 b \gamma,
\end{split}
\end{align}
  the $U(1)$-current,
\begin{align}
J^{gh} = 2bc + 3 \beta \gamma,
\end{align}
and $T^{gh}$ satisfy the relations of the $N=2$ superconformal algebra with the central charge $c_{gh}=-15$.

The ghost fields $b,c$ can be realised in terms of a boson  $\sigma$  with OPE
\begin{align}
\sigma(z)\sigma (0) &\sim \ln z,
\end{align}
by the formulas
\begin{align}
\begin{split}
c &= e^\sigma,\\
b &= e^{-\sigma }.
\end{split}
\end{align}
The ghost fields  $\beta,\gamma$ can be realised by two bosons  $\phi, \chi$ with the OPE's
\begin{align}
\begin{split}
\phi(z) \phi(0)&\sim -\ln z,\\
\chi(z) \chi(0)& \sim \ln z.
\end{split}
\end{align}

The ghosts $\beta,\gamma$ are given by the formulas

\begin{align}
\begin{split}
\beta &= e^{-\phi+\chi}\partial \chi,\\
\gamma &= e^{\phi -\chi} .\\
\end{split}
\end{align}
One can show that these exponents satisfy (\ref{ghost ope}).

The ghost energy-momentum tensor reads
\begin{align}
T_{gh}=T_\phi+T_\chi+T_\sigma,
\end{align}
where
\begin{align}
\begin{split}\label{phi t}
T_\phi  &= - \frac 1 2 \partial \phi \partial \phi - \partial^2 \phi,\\
T_\chi &= \frac 1 2 \partial \chi \partial \chi +\frac 1 2 \partial^2\chi ,\\
T_\sigma &= \frac 1 2 \partial \sigma \partial \sigma +\frac 3 2 \partial^2 \sigma.
\end{split}
\end{align}

The vacua $V_q$ of the $\beta-\gamma$ system in terms of  $\phi$ have the form
\begin{align}
V_q = e^{q \phi}.
\end{align}
By a straightforward computation, one can show that contraction of  $\exp{q\phi}$ with $\beta,\gamma$ has the proper powers of $z$ in the relations (\ref{picture}).
With the use of (\ref{phi t}) one can find the dimension of the vacuum,  $\Delta(\exp(q\phi))= -(q^2+2q)/2$.

Using the formulas
\begin{align}
\begin{split}
\beta \gamma & = \partial \phi,\\
b c &= - \partial \sigma,
\end{split}
\end{align}
one can show that the $U(1)$-current is
\begin{align}
J^{gh}= \partial H^{gh},
\end{align}
where
\begin{align}
H^{gh}= 3 \phi-2\sigma.
\end{align}
\subsection{$N=2$ in NSR string and supercharge}\label{sp time superch}
It was shown in the two last sections that there is  a hidden $N=2$ superconformal symmetry in the ten-dimensional NSR string. The total currents are
\begin{align}
\begin{split}
T^{tot}&= T^{m}+T^{gh},\\
G^{tot}_\pm &=G^m_{\pm}+G^{gh}_\pm,\\
J^{tot}&= J^{m}+J^{gh}.
\end{split}
\end{align}

After bosonization the $U(1)$-current reads
\begin{align}
J^{tot} = \partial H^{tot} = \partial H^{m} + \partial H^{gh} = \sum_k i\partial H_k + 3\partial \phi - 2 \partial \sigma.
\end{align}

 It was discussed in section \ref{n=2}, that the spectral flow can be realised with the use of the vertex operator of the bosonization of the $U(1)$-current. Moreover, the spectral flow with half-integer $\eta$ transform the NS- and R- sectors into each other and is therefore a natural candidate for a the space-time SUSY generator. The spectral flow vertex operator for the total  $N=2$ superconformal  algebra with $\eta = -1/2$ reads
\begin{align}
Q(z) = U_{-1/2} =  \exp \left(-\frac 1 2 H^{tot}\right).
\end{align}
Using the expression of $H^{tot}$ in terms of the bosons, one can rewrite it as
\begin{align}\label{superch}
Q(z) = \exp \left( -\frac 1 2 \sum iH_k -\frac 3 2 \phi + \sigma\right) = c S_\alpha e^{-\frac 3 2 \phi},
\end{align}
where in the last step we returned from the bosons to the original fields. Here $S_\alpha$ is the spin field with all the spins down.

Note, that the vertex operator of the massless fermion in the picture $-3/2$ is
\begin{align}\label{vertex -3/2}
V_{-3/2} = c  u^\alpha S_\alpha e^{-3/2\phi} e^{ikX},
\end{align}
where $ u ^ \alpha $ is a polarization, which must satisfy the Dirac equation, as follows from (\ref{ph st def}). Note that the vertex $ Q (z) $ is the vertex $ V_{- 3/2} (z) $ with  the momentum $ k_ \mu $ equals zero. When the momentum is zero the Dirac equation is trivially satisfied and $ u ^ \alpha $ becomes an arbitrary spinor and can be omitted. This fact is similar to  a supercharge construction in the manuscripts \cite{Friedan:1985ey, Knizhnik:1985ke}, where the supersymmetry  operator  is a fermion vertex operator at zero momentum. In these works it was in the picture $ -1 / 2 $.

The vertex (\ref {vertex -3/2}) is  multiplied by the ghost $ c (z) $ and in the correlation functions it corresponds to the vertex operator with a fixed coordinate. As is usual in string theory, the vertices in the correlation functions must  be multiplied by the ghost  $ c (z) $, and the correlation function does not depend on the coordinate  $ z $, or the vertices must be integrated over	 coordinate $z$. By analogy we replace the ghost $ c (z) $ in (\ref {superch}) with the integral over $ z $, and we understand the action of the supersymmetry operator on the physical vertex as an integral along the contour around it. Accordingly, the supercharge takes the form

\begin{align}\label{supercharge}
Q_\alpha = \int dz \, S_\alpha e^{-3/2\phi}.
\end{align}
By a straightforward computation one can show that it has  dimension 0 and is BRST-closed.

We conclude this section with a couple of comments. First, we got a supercharge with all spins down. By a Lorentz transformation or what is the same by selecting another $ N = 2 $ algebra in the matter sector, we can get the rest of its components. The result is a 16-component Weyl spinor. Second, in the implementation of the spectral flow in section \ref {n=2}, we implicitly use the fact that the central charge does not vanish. For example, as can be seen from (\ref {j boson}), at $ c = 0 $, the $U(1)$-current can not be written as derivative of  a free boson. However, this difficulty is purely technical and can be easily bypassed. For example, consider the spectral flow, built in this section, as the separated  application of the spectral flows of the  matter and ghosts, for which the central charge is not zero. One can also carry out a similar reasoning as in section \ref{n=2}, but bosonize the $ U (1) $-current, not by one, but two bosons $ J ^ {tot} = \partial H ^ m + \partial H ^ {gh} $. As we will do now.

It should be noted  that, unlike the case of $ c \ne 0 $, the exponential  $ \exp (\alpha H ^ {tot}) $ is neutral with respect to the current $ J^{tot} $. Indeed,

\begin{align}
\partial H^{tot} e^{\alpha H^{tot}} = ( 3 \partial \phi -2 \partial \sigma + \sum_k i\partial H_k) e^{\alpha(  3  \phi -2  \sigma + \sum_k iH_k)} = (-9 + 4 + 5 )\alpha e^{\alpha H^{tot}} = 0.
\end{align}
This means, that one cannot make the operator neutral by isolating the $U(1)$-part. This makes our case slightly different and is a consequence of $c=0$.

However, one can proceed as follows. We will make the currents neutral with respect to {\bf total} $U(1)$ by isolating its $U(1)$ ghost and matter components {\bf separately}. For the supercurrents $G^{\pm}$
\begin{align}
G^{\pm}_{tot} = \hat G^{\pm}_m e^{\pm \frac 1 5 H_m} + \hat G^{\pm}_{gh} e^{\mp \frac 1 5 H_{gh}}.
\end{align}
Hereinafter operators with a hat are neutral with respect to the both currents. This formula is a modification of the corresponding formula in section \ref{n=2}. We now consider a field $ V $, which has a charge  $ q_m $ with respect to $ \partial H_m $ and $ q_ {gh} $ with respect to $ \partial H_{gh} $. By isolating charged parts, we get
\begin{align}
V= \hat V e^{\frac {q_m} 5 H_m - \frac {q_{gh}} 5 H_{gh}}.
\end{align}
This operator has a charge $ q_m + q_ {gh} $ with respect to $ \partial H ^ {tot} $. The signs in front of the exponents $ H ^ m $ and $ H ^ {gh} $ are different because the central charges $ c_m = $ 15 and $ c_ {gh} = - $ 15 have  different signs. Now let's make the twist with the help of the operator $ \exp (\eta H) $

\begin{align}
V_\eta = \hat V \exp\left\{\frac 1 5 (5\eta + q_m)H_m+\frac 1 5 (5\eta - q_{gh})H_{gh}\right\}.
\end{align}
Considering the operator expansion with $ G ^ {\pm} $, it is easy to see that it will be expanded in powers of $ (z-w) ^ {n \pm \eta} $. It is also clear that the full charge of the field does not depend on $ \eta $
\begin{align}
q' = q = q_m+q_{gh},
\end{align}
since, as we have shown, $ \exp (\eta H ^ {tot}) $ is neutral with respect to the $ U (1) $-current $ \partial H ^ {tot} $. The conformal dimension of  $V_\eta$ is
\begin{align}
\Delta' = \Delta  - \frac 1{10} (5\eta - q_{gh})^2 + \frac 1{10} (5\eta - q_{m})^2 + \frac 1 {10}q_{gh}^2 -  \frac 1 {10}q_{m}^2 = \Delta +  \eta q.
\end{align}
 All of these calculations, show that the operator $ \exp (\eta H ^ {tot}) $ produces the twist (\ref {automorph}) with $ c = 0 $.
\section{The GSO-projection and the action of the supercharge}\label{section gso}
In the previous section we constructed the operator $ Q_ \alpha $, which has the conformal dimension 0 and commute with the BRST-operator $ Q_B $. However, this is not enough to act on the physical (BRST-closed) vertices. It is necessary to retain only such physical vertices for which  the action of $ Q_\alpha $ is correctly defined. Consider an arbitrary vertex
\begin{align}\label{ph ver}
P(\partial X_\mu, \partial H_k,\partial \sigma,\partial \phi,\partial \chi )\exp\left[l\phi +r \chi +m \sigma+  \sum_k i s_k H_k + i k_\mu X^\mu\right],
\end{align}
where $ P $ - a polynomial in its arguments. In what follows its exact form is not important.

Recall, that in the  R-sector $l,s_k\in \mathbb Z+\frac 1 2$, and in the NS-sector $l,s_k\in \mathbb Z$. In both sectors $r,m \in \mathbb Z$.

When translating the integrand in (\ref{supercharge}) around the vertex a phase occurs
\begin{align}
2\pi i\left(\frac{3l}2  - \frac 1 2 \sum s_k \right).
\end{align}
In order to avoid any cuts, and for the integral in the action of the supercharge, to be well defined,  the physical vertices (\ref {ph ver}) must be mutually local with the supercurrent (\ref {supercharge})
\begin{align}\label{gso}
\frac {3l}2- \frac 1 2 \sum s_k \in \mathbb Z.
\end{align}
This requirement leaves  only those vertices that satisfy condition (\ref {gso}) in each sector.

We now show that this condition is equivalent to the GSO-projection. Recall that the GSO-projection by definition \cite{Polchinski:1998rq} leaves in the spectrum of  the string only states with  eigenvalues 1 of the operator $ (- 1) ^ F $, where $ F $ is the fermion number. By definition, the fermion number $ F_ {gh} $ of  the exponential $ e ^ {l \phi} $ is equal to $ l $. While in the matter sector the fermion number is defined as
\begin{align}
F_m = \int dz\, \sum_k \Sigma_{2k,2k+1}(z) = \int dz   \sum_k \psi_{2k}\psi_{2k+1} = \int dz \sum_k {i\partial H_k},
\end{align}
where $\Sigma_{\mu\nu}$ is the fermionic part of the Lorentz transformation. Full fermion number is equal to their sum $F=F_{m}+F_{gh}$.

The action on vertex (\ref {ph ver}) of the operator $ F = F_ {gh} + F_ {m} $ will give
\begin{align}
l + \sum_k s_k.
\end{align}
The action of the operator $(-1)^F$ on a physical vertex is equal  to 1, when
\begin{align}
l + \sum_k s_k \in 2\mathbb Z.
\end{align}
This condition is equivalent to (\ref{gso}). Indeed, if one divided it by 2 and add to (\ref{gso}), then
\begin{align}
2 l  \in \mathbb Z,
\end{align}
which is true since  $l$ is at least half-integer.

From the equivalence of (\ref {gso}) and the GSO-projection follows a standard assertion of the equality of the number of the physical states of bosons and fermions.  This is equivalent to the vanishing of the one-loop correction to the energy of the vacuum.

We give a few examples of the action of projection (\ref {gso}) on the physical states. The vertex of a tachyon

\begin{align}
V = e^{ikX}e^{-\phi},\quad\quad  k^2= \frac 1 2,
\end{align}
 for it the left-hand side of the expression (\ref {gso}) is $ 3/2 $. Since $ 3/2 $ is not an integer, the tachyon is absent from the spectrum after the projection.

The vertex of the massless fermion is
\begin{align}
V=u^\alpha S_\alpha e^{ikX}e^{-\phi/2} = u^{\alpha}e^{\sum is_k H_k}e^{ikX}e^{-\phi/2} ,\quad\quad  k^2= 0,
\end{align}
where  $s_k=\pm 1/2$ corresponds to polarization of spinor $\alpha$.
Condition  (\ref{gso}) for this vertex is
\begin{align}
-\frac 3 4 -\frac 1 2 \sum_k s_k \in\mathbb Z.
\end{align}
It means that an odd number of spins $ s_k $ should be directed upwards. This leaves the 16-component Weyl spinor out of the 32-component Dirac spinor.

The vertex operator of the massless boson is
\begin{align}\label{boson gso}
V= \xi_\mu \psi^\mu e^{ikX}e^{-\phi},\quad\quad  k^2= 0.
\end{align}
For a term  $ \xi_ \mu \psi ^ \mu $ the sum of $ \sum_k s_k $ is $ \pm $ 1 (or more precisely, it is divided into the sum of two components with charge $ 1 $ and $ -1 $). Accordingly, condition (\ref {gso}) takes the form
\begin{align}
-\frac 3 2 \pm \frac 1 2 \in \mathbb Z.
\end{align}
It is true for both signs. Therefore, all physical vertices of the form (\ref {boson gso}) are present in the spectrum after the projection.

To summarize, there are vector boson and Weyl fermion on the massless level after the projection.

Thus, the requirement of the absence of cuts in the action of the operator $ Q_ \alpha $ on the physical states leads to the GSO-projection of this space. From this it is clear that the GSO-projection is necessary for the very possibility  of determining the space-time supersymmetry action on the physical states.
\section{The massless states at $d=10$}
\subsection{The vertex operators}
The vertex operator in the NS-sector in the picture $-1$ is
\begin{align}
V_{-1} = c \xi_\mu \psi^\mu e^{ikX}e^{-\phi},
\end{align}
where $\xi_\mu$ is polarization.

Note, that the matter part $\psi_\mu$, as well as the ghost part\footnote{The ghost sector is not unitary. Therefore the relations (\ref{ramond ineq}) and (\ref{chiral ineq}) do not work in this case. We will use (\ref{chiral mode def}) and (\ref{ramond mod def}) as the definition of the chiral fields and the ramond vacua } $c e^{-\phi}$ are the chiral fields of the  $N=2$ superconformal algebra.

On the other hand, the vertex of the massless fermion in the picture $-1/2$ is
\begin{align}
V_{-1/2} = сu^\alpha S_\alpha e^{ikX} e^{-\phi/2}.
\end{align}
Note, that the matter part $S_\alpha$, has the dimension $5/8$ and is therefore Ramond vacuum,  as well as $ce^{-\phi/2}$ in the ghost part.

So, as mentioned above,  the vertices of the massless particles are constructed from the chiral fields and  the Ramond  vacua. They are converted into each other under the action  of the space-time supersymmetry, forming a supermultiplet. The same remains true after compactification.

\subsection{The action of supercharge}
It follows from (\ref{ch -> ram}), that the action of the spectral flow interchange the chiral fields and the Ramond vacua, which correspond to  massless bosons and fermions. As we have shown in section \ref{sp time superch}, the space-time supersymmetry operator is precisely the spectral flow. Formulas (\ref{ch -> ram}) show that the action of supercharge transforms bosons and fermions into each other. In this section we will write the supersymmetry action explicitly.

For this purpose, it is convenient to write the supercharge and the vertices in the different pictures. We will give these formulas without derivation. The supercharge
\begin{align}
\begin{split}
Q^{(1/2)}_\alpha&=\int dz\,\, e^{\phi/2} (\gamma_\mu)_{\alpha\beta} S'_\beta \partial X_\mu ,\\
Q^{(-1/2)}_\alpha&=\int dz\,\,e^{-\phi/2}S_\alpha,\\
Q^{(-3/2)}_\alpha&=\int dz\,\,e^{-3\phi/2}S'_\alpha.
\end{split}
\end{align}
Hereinafter $S_\alpha$ is a right Weyl spinor, $S'^\beta$ is left.

The vertex operators of the boson is
\begin{align}
\begin{split}
V_0 &= \xi_\mu \left(\partial X^\mu + i (k\psi) \phi^\mu\right) e^{ikX},\\
V_{-1}&= \xi_\mu \psi^\mu e^{ikX}.
\end{split}
\end{align}
The fermionic vertex is
\begin{align}
\begin{split}
V_{1/2} &= u^\alpha(\gamma^\mu)_{\alpha\beta} S'_\beta (\partial X + \frac i 2 (k \psi)\psi_\mu)e^{ikX}e^{\phi/2},\\
V_{-1/2} &= u^\alpha S_\alpha e^{ikX} e^{-\phi/2},\\
V_{-3/2} &= u^\alpha S'_\alpha e^{ikX} e^{-3\phi/2}.
\end{split}
\end{align}
Using these formulas one can show that up to numerical coefficients the following holds
\begin{align}
\begin{split}
[v^\alpha Q_\alpha,V_{\text{NS}}(\xi,k)] &= V_{\text{R}}(k^\mu \xi^\nu {(\gamma_{\mu\nu})^\alpha}_\beta v^\beta,k),\\
[v^\alpha Q_\alpha,V_{\text{R}}(u,k)] &= V_{\text{NS}}(u^{\alpha}(\gamma^\mu)_{\alpha\beta}v^\beta,k),
\end{split}
\end{align}
where we dropped the picture numbers since it holds for all the  pictures.
\section{Compactification}
 In this section we will briefly discuss the compactification of six dimensions. In order to preserve the $ N = 1 $ space-time supersymmetry after superstring compactification, it is necessary and sufficient \cite{Gepner:1987qi,Banks:1987cy}, to preserve the $ N = 2 $ superconformal symmetry on the world-sheet. Since the ghost and a free space-time part have the $ N = 2 $ superconformal symmetry, it is necessary to possess an $ N = 2 $ superconformal symmetry in the compact part. The supercharge is constructed as the spectral flow in the three components. The ghost part of the $N=2$ superconfromal algebra is the same as it was in the ten-dimensional case. The space-time part is similar to that of the ten-dimensional case, but is built from 4 $\psi_\mu, X_\mu$. Namely,
   \begin{align}
   \begin{split}
   G_{\pm}^{\text{sp-t}}&= i\psi_0^\pm \partial X^{\mp}_0  +i\psi_1^\pm \partial X^{\mp}_1,\\
   J^{\text{sp-t}} &= \psi_0^+\psi_0^-+\psi_1^+\psi_1^-= i\partial H_0 + i\partial H_1.
   \end{split}
   \end{align}
   This part has the central charge $c=6$.

  As we have discussed, the compact part must have the $N=2$ superconformal symmetry. We will only need the $U(1)$-current, which bosonizes as
  \begin{align}
  \begin{split}
  J^{int} &= i\partial H^{int},\\
  H^{int}(z)H^{int}(0)&\sim -3 \ln z.
  \end{split}
  \end{align}
  As well as in the ten-dimensional case, the supercharge can be build as a spectral flow of the full $N=2$ superconfromal algebra with the parameter $\eta=-\frac 1 2$
  \begin{align}
 Q (z)= \exp \left [- \frac 1 2 H ^ {tot} \right] = \exp \left [- \frac 1 2 \left (H ^ {gh} + H ^ {\text {sp-t}} + H ^ {int} \right) \right].
  \end{align}
  We can rewrite it as
  \begin{align}
  Q(z) = \exp \left [-\frac 3 2 \phi + \sigma  - \frac 1 2 iH_0 - \frac 1 2 i H_1 - \frac 1 2 i H^{int} \right] = c S_\alpha \Sigma e^{-\frac 3 2 \sigma},
  \end{align}
  where we used the explicit expressions for $H ^ {gh}$, $H ^ {\text {sp-t}}$ and returned to the original fields. $S_\alpha$ is a four-dimensional spinor built similarly to (\ref{spinors}) and we used the notation $\Sigma = \exp (-\frac 1 2 i H^{int})$. The polarization $\alpha$ corresponds to all spins down $s_0=s_1 = -\frac 1 2 $. By the Lorentz transformation we can get another one $s_0=s_1 = \frac 1 2$. These two polarization form 2-component Weyl spinor in four dimensions. Replacing the ghost $c$ by an integral, we achieve
  \begin{align}\label{supercharge comp}
  Q= \int dz \, S_\alpha \Sigma e^{-\frac 3 2 \sigma}
  \end{align}
  Reasoning similar to the one in section \ref{section gso} leads to the restriction on the physical vertices in order to have well-defined action of the supercharge. The locality condition reads
  \begin{align}\label{gso comp}
  \frac {3l}2- \frac 1 2 s_0 - \frac 12  s_1 - \frac 12 Q^{int} \in \mathbb Z,
  \end{align}
where $Q^{int}$ is charge of the physical vertex with respect to the current $ J^{int}$.

It is instructive to study the massless spectrum of the compactified string. The massless bosons can be a vector or a scalar. The vertex operator for the massless vector boson reads
\begin{align}\label{vector}
V = c \xi^\mu\psi_\mu e^{-\phi}e^{ikX}.
\end{align}
The space-time and the ghost parts of it are chiral fields of the corresponding $N=2$ superconformal algebras.

The scalar boson vertex operator must have the form
\begin{align}\label{scalar}
V = c \Lambda e^{-\phi}e^{ikX},
\end{align}
where $\Lambda$ is some scalar field of the compact part with the dimension $\frac 1 2 $. The inequality $|Q^{int}|\le 2 {\Delta^{int}} $ and condition (\ref{gso comp}) leads to two possibilities for the charge  $Q^{int} = \pm 1 $. This means that $\Lambda$ must be a chiral field of the compact theory. The opposite is also true. One can show that for any chiral field $\Lambda$ of the compact part with dimension $\frac 1 2 $ vertex (\ref{scalar}) is BRST-closed. The property of  chiral primary field $G^{-}_{\frac 1 2 } \Lambda=0$ (or $G^{+}_{\frac 1 2 } \Lambda=0$ for antichiral primary field) is crucial for this.

The fermion vertex operator must be a space-time spinor and therefore it has the form
\begin{align}\label{fermion comp}
V =c u^\alpha S_\alpha \Omega e^{-\phi/2 }e^{ikX},
\end{align}
where $\Omega$ is a field of the compact part and $\alpha$ corresponds to $s_0=s_1 = \pm \frac 12$. In order for $V$ to have conformal dimension zero, $\Omega$ must have dimension $\frac 3 8$. Since $\frac 3 8 = \frac {c_{int}} {24}$ field $\Omega$ must be the Ramond vacuum of the compact part. $S_\alpha$ and $ce^{-\phi/2 }$ are the Ramond vacua of the space-time and the ghost parts as well. Vertex (\ref{fermion comp}) is a two component right Weyl spinor. The left Weyl spinor\footnote{There is also left Weyl part of supercharge (\ref{supercharge comp}).} vertex are
\begin{align}
V= c v^\alpha S'_\alpha \overline \Omega e^{-\phi/2 }e^{ikX},
\end{align}
where $\overline \Omega$ is an image of $\Omega$ under the automorphism $J^{int} \rightarrow -J^{int}$ of the  $N=2$ algebra and $\alpha$ corresponds to $s_0 = -s_1=\pm \frac 1 2$. The condition (\ref{gso comp}) and the inequality for dimension $\Delta^{int} \ge \frac {{(Q^{int})}^2} {6}$  restrict the possible charge $Q^{int}$ of $\Omega$ to be $\frac 1 2 $ or $-\frac 3 2$ ($-\frac 1 2 $ or $\frac 3 2 $ for $\overline \Omega$). The opposite is also true. For any Ramond vacuum of the compact part $\Omega$ with charge $\frac 1 2 $ or $-\frac 3 2$ vertex (\ref{fermion comp}) is BRST-closed. The property of the ramond vacuum $G_0 \Omega =0$ is again crucial for this. Note, that there are only one field $\Sigma = \exp({-\frac 1 2 H^{int}})$ with charge $\frac 3 2 $ and its vertex is superpartner of the unique  vector boson (\ref{vector}) of theory. The action of supercharge can be derived from the formulas (\ref{ch -> ram}).

It is worth noting that the proposal about connection between the $N=2$ superconformal symmetry and the space-time supersymmetry of the compactified string models also was suggested in \cite{Sen:1986mg,Sen:1986pw}.
\section{Conclusion}
It is shown that the hidden  $ N = 2 $ superconformal symmetry, which takes place in ten-dimensional NSR string in the matter and the ghost sectors, allows to obtain the space-time supersymmetry  in the similar way as  it was done by Gepner in compactification of six dimensions. Note that we used the covariant quantization of the string meanwhile in \cite{Gepner:1987qi} the light-cone gauge was used.

The spectral flow of the $ N = 2 $ superconformal algebra allows us to construct a bijection between boson (NS) and fermion (R) sectors subject to the limitation of space of the physical states on a subspace of  the states which are local with respect to  the spectral flow operator. This reduction is equivalent to the GSO-projetion.

The vertices of the massless particles correspond to the chiral fields with  the $ U (1) $ - charge equal to $\pm1$ in the case of bosons and to the Ramond vacua in the case of fermions.

The action of the spectral flow combines vertices of the  massless particles in one supermultiplet.
\\

We are grateful to S. Parkhomenko, D. Polyakov and G. Tarnopolsky for usefull comments and discussions. Work of L. Spodyneiko was supported by Russian Foundation for Basic Research under the grant RBRF 	15-32-20974.
\bibliographystyle{unsrt}
\bibliography{spodyneiko}
\end{document}